\documentclass[fleqn,usenatbib]{mnras}

\usepackage{newtxtext,newtxmath}
\usepackage[T1]{fontenc}
\usepackage{ae,aecompl}

\usepackage{graphicx}	% Including figure files
\usepackage{amsmath}	% Advanced maths commands
\usepackage{amssymb}	% Extra maths symbols
\usepackage{url}

\newcommand{\Fermi}{\textit{Fermi}}
\newcommand{\Gaia}{\textit{Gaia}}
\def\ltsima{$\; \buildrel < \over \sim \;$}
\def\simlt{\lower.5ex\hbox{\ltsima}}
\def\gtsima{$\; \buildrel > \over \sim \;$}
\def\simgt{\lower.5ex\hbox{\gtsima}}

\title[A hunt for dSphs towards \Fermi-LAT sources]{A \Gaia\ DR2 search for dwarf galaxies towards \Fermi-LAT sources: implications for annihilating dark matter}

\author[I. Ciuc\u{a} et al.]
{\parbox{\textwidth}{Ioana Ciuc\u{a}$^{1, 2}$\thanks{E-mail: ioana.ciuca.16@ucl.ac.uk}, Daisuke Kawata$^{1}$,
		Shin'ichiro Ando$^{3,4}$,
        Francesca Calore$^{5}$,
		Justin I. Read$^{6}$ and
        Cecilia Mateu$^{7}$}\vspace{0.5cm}
	\\
	$^{1}$ Mullard Space Science Laboratory, University         College London,
	Holmbury St. Mary, Dorking, Surrey, RH5 6NT, UK \\
    $^{2}$ LSSTC Data Science Fellow\\
	$^{3}$ GRAPPA Institute, University of Amsterdam, 1098 XH Amsterdam, Netherlands \\
    $^{4}$ Kavli Institute for the Physics and Mathematics of the Universe, University of Tokyo, Kashiwa 277-8583, Japan \\
    $^{5}$ Univ. Grenoble Alpes, USMB, CNRS, LAPTh, F-74940 Annecy, France \\
    $^{6}$ Department of Physics, University of Surrey, Guildford, GU2 7XH, UK \\
    $^{7}$ Centro de Investigaciones de Astronom\'ia, AP 264, M\'erida 5101-A, Venezuela \\
}

\date{Accepted 2018 July 16. Received 2018 June 29; in original form 2018 May 15}

\pubyear{2018}

\begin{document}
\label{firstpage}
\pagerange{\pageref{firstpage}--\pageref{lastpage}}
\maketitle

\begin{abstract}
We make the first attempt to find dwarf galaxies in eight \Fermi-LAT extended, unassociated, source fields using \Gaia\ DR2. After probing previously unexplored heliocentric distances of $d<20$~kpc with an extreme-deconvolution (XD) technique, we find no sign of a dwarf galaxy in any of these fields despite \Gaia's excellent astrometric accuracy. Our detection limits are estimated by applying the XD method to mock data, obtaining a conservative limit on the stellar mass of $M_* < 10^4$~M$_{\sun}$ for $d < 20$\, kpc. Such a low stellar mass implies either a low-mass subhalo or a massive stripped-down subhalo. We use an analytic model for stripped subhalos to argue that, given the sizes and fluxes of the \Fermi-LAT sources, we can reject the hypothesis that they owe to dark matter annihilation.
\end{abstract}


%%%%%%%%%%%%%%%%%%%%%%%%%%%%%%%%%%%%%%%%%%%%%%%%%%

%%%%%%%%%%%%%%%%% BODY OF PAPER %%%%%%%%%%%%%%%%%%

\section{Introduction}
\label{sec:intro}

In the current $\Lambda$CDM cosmological model, dark matter (DM) is the dominant matter component of the Universe, comprising approximately 26.8\% of its total mass-energy \citep[e.g.][]{Planck+XIII16}. It is required to explain the rotation curves of stars and gas in galaxies \citep[e.g.][]{1980ApJ...238..471R,2011AJ....141..193O,2013NewAR..57...52B}, strong and weak lensing of galaxies and clusters \citep[e.g.][]{2006ApJ...648L.109C,2015Sci...347.1462H}, and the growth of large scale structure \citep[e.g.][]{2006Natur.440.1137S,2016JCAP...08..012B}, yet its nature remains elusive.

The latest observational evidence points to DM being a cold, collisionless, fluid \citep[e.g.][]{2006ApJ...648L.109C,2006PhRvL..96a1301S,2011IJMPD..20.2749D,2016JCAP...08..012B,2017MNRAS.467.2019R}, indicative of a new, fundamental, particle not present in the Standard Model of particle physics \citep[e.g.][]{2005PhR...405..279B}. Of the many candidate particles, one of the most well-motivated is a Weakly Interacting Massive Particle (WIMP), with a mass in the range $2 \simlt m_\chi/{\rm GeV} \simlt 2000$\,GeV \citep[e.g.][]{1996PhR...267..195J}. Depending on its mass and annihilation cross section, WIMP DM can annihilate or decay into Standard Model particles, including gamma rays \citep[e.g.][]{1987ApJ...313L..47S}. This theorized property of DM has prompted the search for gamma-ray signals from astronomical targets with high DM content as an indirect detection method, using data from space- and ground-based gamma-ray telescopes, such as the Large Area Telescope (LAT), aboard the \Fermi\ satellite, which maps the whole sky from about 30 MeV to $>$500 GeV \citep[e.g.][]{2009ApJ...697.1071A}, and the planned CTA array \citep[e.g.][]{2013APh....43....3A}.

Dwarf spheroidal galaxies (dSphs) in the Milky Way represent a particularly clean target for gamma-ray emission due to DM annihilation because of their high mass-to-light ratios~\citep[e.g.][]{Aaronson83,Pryor+Kormendy90,Mateo98,2011MNRAS.418.1526C,2015MNRAS.453..849B,2017PhRvD..95l3012K} and relatively low gamma-ray background emission from astrophysical sources \citep[e.g.][]{1990Natur.346...39L,Evans+04,Baltz+08,Winter:2016wmy}. So far, the search for a DM gamma-ray annihilation signal from such systems has not yielded any significant detection in the frequency range probed by \Fermi-LAT. However, it provides some of the most stringent constraints on the nature of DM particles in terms of their mass and annihilation cross-section~\citep[e.g.][]{2017ApJ...834..110A,2018arXiv180305508C}. This is beginning to challenge the WIMP paradigm.

Usually, dSphs are identified in optical surveys and confirmed via deep photometric or spectroscopic follow-up \citep[e.g.][]{2007ApJ...654..897B}. Then, a corresponding gamma-ray signal can be looked for in their direction \citep[e.g.][]{2015PhRvL.115h1101G,2017ApJ...834..110A,2018arXiv180305508C}. An alternative approach is to explore the association of stellar counterparts with ``unassociated" gamma-ray sources, i.e. sources identified as such by \Fermi-LAT but lacking counterparts at other wavelengths. The detection of a dSph at the position of a gamma-ray source would provide the first observational evidence of the association between a dSph and gamma-ray emission. If the gamma-ray source is spatially extended, then this would be a `smoking gun' for DM annihilation \citep[e.g.][]{Bertoni+15,Bertoni+16,2017PhRvD..96f3009C}.

Unassociated sources represent almost one third of all gamma-ray detected sources in the Third \Fermi~Gamma-Ray Source Catalog \citep[the 3FGL][]{Acreco+Fermi+15}. Among 3FGL unassociated sources, \citet{Bertoni+15,Bertoni+16} found that 3FGL J2212.5+0703 exhibits a spatially extended profile, with no other wavelength counterparts so far, and suggested that this could be a DM subhalo.  \citet{Xia+17} found that the source 3FGL J1924.8-1034 also has a spatially extended profile at high significance, making it another possible DM subhalo candidate. We caution, however, that~\cite{2018arXiv180408035F} did not find any evidence for an
individual source with statistically significant extension at the position of these two objects; a model for two close point-like sources was preferred in both cases.
Very recently, the \Fermi-LAT Collaboration released the catalogue of extended high-latitude sources, $|b|>5\degr$~\citep{2018arXiv180408035F}, where six newly unassociated extended objects are identified.

The European Space Agency's \Gaia\ mission \citep{Gaia+Prusti+16} has made the second data release of their unprecedented parallax and proper motion measurements for more than one billion stars brighter than $G<20$ mag \citep[\Gaia\ DR2;][]{Gaia+Brown+18,Lindegren+18}. This provides us with a new window to find dSphs using proper motions in the inner Galactic halo, where the stellar density is too high to detect them from the photometric data alone \citep{Antoja+15}. In this paper, using both parallax and proper motion data from \Gaia\ DR2, we search for dSphs in the fields of the two above mentioned \Fermi-LAT sources as our primary target. Additionally, we apply the same technique to the six \Fermi-LAT extended source fields recently found by \citet{2018arXiv180408035F}.

We focus on the (heliocentric) distance range between 1 and 20\,kpc. So far, no dSph galaxies have been found nearer than 20\,kpc from the Sun \citep{Belokurov14,Koposov15,Drlica-Wagner15,Bechtol15,Kim15,KimJerjen15,Laevens15a,Laevens15b,Luque16}. Only Draco II is estimated to be at a distance of 20\,kpc \citep{Laevens15b,2016MNRAS.458L..59M}, but the vast majority ($\gtrsim70$\%) of dSphs have been found at distances farther than 50\,kpc. This owes to a combination of several effects. In the inner halo, dSph satellites are expected to be more diffuse both \emph{intrinsically} and \emph{apparently}: intrinsically because they are more prone to tidal disruption induced by the Galactic disc \citep[e.g.][]{2010ApJ...709.1138D,Garrison-Kimmel+17,Sawala+17}; and apparently, because of the projected size which increases with decreasing distance. At larger apparent sizes, the contamination by field halo and disc stars plays an increasingly important role, washing out a dSph's signature on the sky. Thus, the inner halo within 20 kpc from the Sun is an unexplored territory. \Gaia\ DR2 can make a significant contribution over this distance range by making use of joint distance and kinematic information to find both phase space overdensities, and to separate dSph stars from the Milky Way foreground \citep{Antoja+15}. By combining our constraints on the presence or absence of dSphs at these small distances with the current lack of a DM annihilation signal from the more distant dSphs \citep{2017ApJ...834..110A}, we test the hypothesis that these unassociated \Fermi-LAT sources owe to DM annihilation.

This paper is organised as follows. Section~\ref{sec:meth} describes the method to search for stellar counterparts to the \Fermi-LAT unassociated, extended sources. In Section \ref{sec:res} we present the results of our search for a dSph within \Gaia\ DR2. In Section \ref{sec:mock}, we discuss the detection limits of our method with \Gaia\ DR2, and how the upper bound on $M_*$ translates into bounds on the likely pre-infall halo mass of a dSph. In Section~\ref{sec:dm-disc}, we use our detection limits from Section \ref{sec:mock} to determine whether or not the \Fermi-LAT sources could be explained by DM annihilation from a nearby subhalo. We find that the sizes and fluxes are inconsistent with the DM subhalo hypothesis. Finally, a summary and discussion of our results are presented in Section~\ref{sec:sum}.

\section{Data and Method}
\label{sec:meth}

We selected a region of the sky within 2$\degr$ and 1$\degr$ from our two primary targets of \Fermi-LAT unassociated gamma-ray sources: 3FGL J2212.5+0703 ($l = 68\fdg74$, $b = -38\fdg57$) and 3FGL J1924.8+1034($l = 27\fdg16$, $b = -12\fdg17$), respectively. We also applied the same technique within 1$\degr$ to the six extended unassociated gamma-ray sources in \citet{2018arXiv180408035F}, namely FHES J1501.0$-$6310 ($l=316\fdg95$, $b=-3\fdg89$), FHES J1723.5$-$0501 ($l=17\fdg90$, $b=16\fdg96$), FHES J1741.6$-$3917 ($l=350\fdg73$, $-4\fdg72$), FHES J2129.9+5833 ($l=99\fdg13$, $b=5\fdg33$), FHES J2208.4+6443 ($l=106\fdg62$, $b=7\fdg15$) and FHES J2304.0+5406 ($l=107\fdg50$, $b=-5\fdg52$). 

We downloaded all of the \Gaia\ DR 2 stars in these regions from the \Gaia\ archive site (\url{https://gea.esac.esa.int/archive/}) and Centre de Donn{\'e}es astronomiques de Strasbourg (CDS). We employed two different strategies for finding a dSph galaxy depending on whether it is located at a distance farther than $d=10$\,kpc or not. 

We used both spatial and proper motion information from \Gaia\ DR2 to determine whether we can observe the stellar imprint of a dSph galaxy at the position of each \Fermi-LAT unassociated, extended,  gamma-ray source. For this purpose, we selected the stars around a radius of 2$\degr$ and 1$\degr$ centred on the position of 3FGL J2212.5+0703 and 3FGL J1924.8+1034, respectively. The smaller field of view for 3FGL J1924.8+1034 was chosen to reduce the contamination from the field stars at low Galactic latitude. The additional 6 fields in \citet{2018arXiv180408035F} are also low Galactic latitude fields, and therefore we chose 1$\degr$ field radius for these sources as well.

When searching for a dSph galaxy farther than $d=10$\,kpc away, we applied the same parallax filter used in \citet{Antoja+15}, namely, we discard stars for which $\varpi - \sigma_{\rm \varpi} > 0.1$\,mas, where $\varpi$ represents the parallax and $\sigma_{\rm \varpi}$ is the parallax uncertainty. This filter corresponds to eliminating stars located at a distance of less than 10\,kpc within parallax uncertainties and aims to minimize contamination from foreground stars. After these quality cuts, we are left with 17,747 and 125,891 stars in the fields of 3FGL J2212.5+0703 and 3FGL J1924.8+1034, respectively. For the additional 6 fields, we obtain 467,082, 61,091, 670,424, 120,156, 82,638 and 115,848 stars in the fields of FHES J1501.0$-$6310, FHES J1723.5$-$0501, FHES J1741.6$-$3917, FHES J2129.9+5833, FHES J2208.4+6443 and FHES J2304.0+5406, respectively.

We then used an Extreme-Deconvolution \citep[XD,][]{Bovy+11} Gaussian Mixture Model \citep[XDGMM,][]{Holoien+17} to perform density estimation on a four-dimensional dataset comprised of the stellar position in Galactic longitude and latitude, and the RA and DEC proper motion measurements. Our aim is to detect a group of stars with similar proper motions which are also concentrated on the sky. As XD has been proven to be a powerful tool in estimating the noise-free underlying distributions of astrophysically relevant quantities \citep{Hogg+05, Bovy+09ApJ...700.1794B, Bovy+12ApJ...749...41B}, we have decided to use it for the current purpose. XD allows us to assume Gaussian errors in the \Gaia\ DR2 proper motions and parallaxes \citep{Hogg+05, Bovy+09ApJ...700.1794B, Bovy+12ApJ...749...41B} and also to take into account the correlation between the measurement of RA and DEC proper motions. We do not take into account the measurement uncertainties in the stellar position or correlations between the positions and proper motions, because the uncertainties in the position in the \Gaia\ DR2 are very small. For this analysis, we applied a small constant uncertainty of $0\fdg01$ in the stellar positions for practical reasons, whose effect is small enough not to affect our results.

\begin{figure}
	\centering
	\includegraphics[width=\columnwidth]{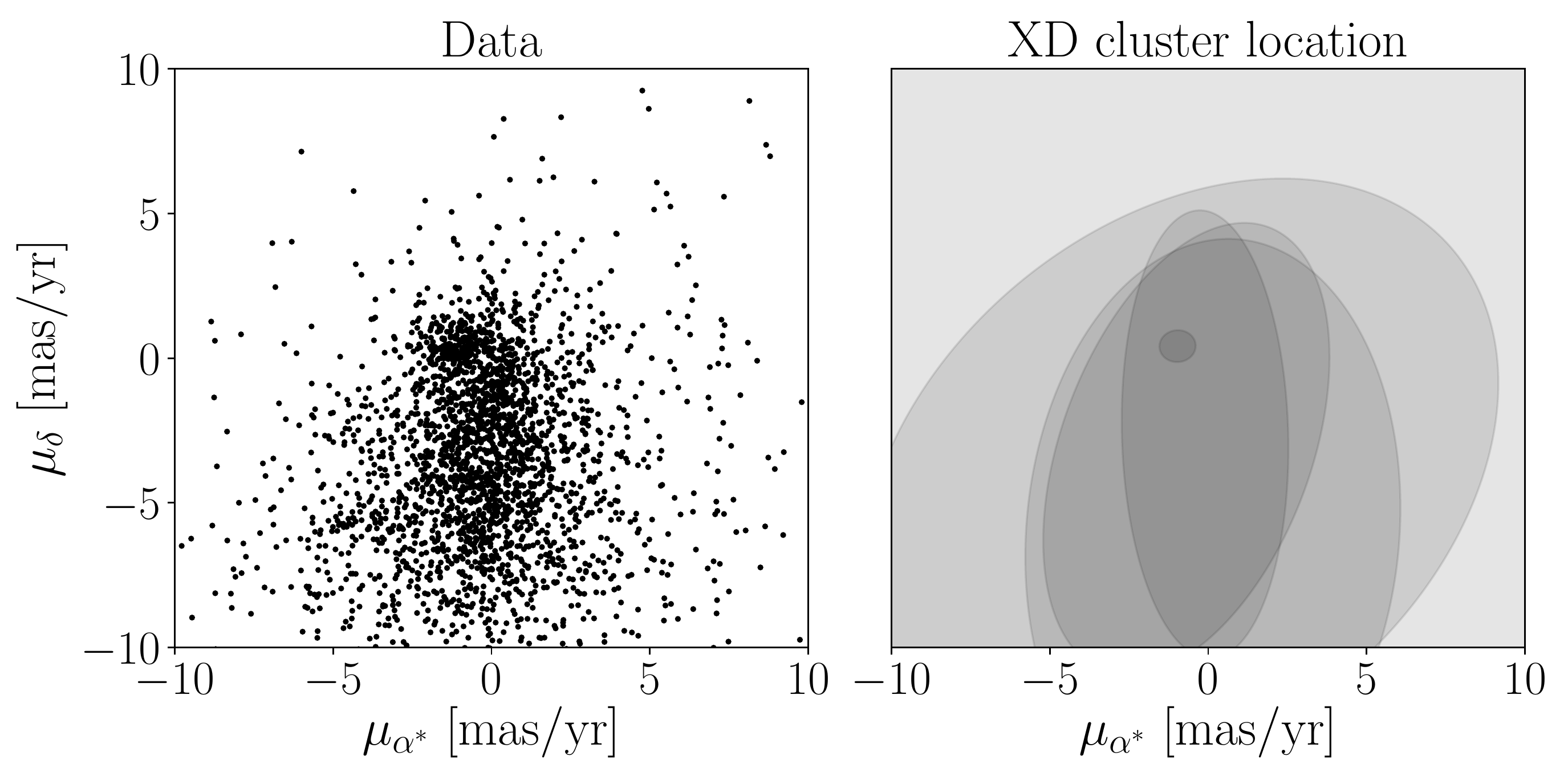}
	\caption{Example of a clear detection of a cluster associated with a dSph galaxy obtained by applying the XD method to mock data. We assumed a dSph galaxy with a stellar mass of $M_*=5000$\,M$_{\sun}$, a half-mass radius of $r_{\rm h}=50$\,pc, and moving at a velocity of $V_{\rm gal}=50$\,km\,s$^{-1}$ at a distance of $d=10$\,kpc in the search field of 3FGL J2212.5+0703. We used $N=6$ components in the XD. The left panel shows the the input stellar distribution in proper motion space, compressed by a factor of 5. The right panel shows the Gaussian components found by XD, including the cluster associated with the mock dSph galaxy centered at a proper motion of $(\mu_{\rm \alpha^*},\mu_{\delta}) \approx (-1,2)$~mas yr$^{-1}$.} 
	\label{fig:4dxd}
\end{figure}

Fig.~\ref{fig:4dxd} shows an example of the XD method applied to a dSph galaxy of stellar mass of $M_*=5000$~M$_{\sun}$, half-mass radius of $r_{\rm h}=50$~pc, and moving at a velocity of $V_{\rm gal}=50$~km~s$^{-1}$ at a distance of $d=10$~kpc in the search field of 3FGL J2212.5+0703. We constructed the background stars using {\tt Galaxia} \citep{Sharma+11ApJ...730....3S} in the field of 3FGL J2212.5+0703. The stellar catalogue, representative of a dSph galaxy, was made with {\tt SNAPDRAGONS} \citep{Hunt+15MNRAS.450.2132H} assuming an age of 12 Gyr, a metallicity of $Z/Z_{\sun} = 0.01$, and a Salpeter initial mass function. The dSph galaxy's dynamical properties were modelled following a simple Plummer distribution with $r_{\rm h}=50$~pc, and an isotropic velocity distribution with a dispersion of $\sigma=10$\,km/s. We applied the expected \Gaia\ DR2 uncertainties at \url{https://www.cosmos.esa.int/web/gaia/dr2} 
to the data for both the dSph galaxy and field stars. 
In Fig.~\ref{fig:4dxd}, we see that XD can find the modelled dSph galaxy in proper motion space. We found that an excessive number of components in the XD does not penalise the detection of the dSph \citep[see also][]{Anderson+2017arXiv170605055A}. However, it may result in overfitting the data. We find that using $K=25$ components is a good model for most of the fields. However, for the low Galactic latitude fields, we additionally used $K=50$ for the fields of 3FGL J1924.8+1034, FHES J1501.0$-$6310 and FHES J2304.0+5406 and $K=100$ for the field of FHES J1741.6$-$3917. The numbers of components for each field are determined such as to recover a dSph with our conservative stellar mass limit, as discussed in Section~\ref{sec:mock}.

For the search for a dSph galaxy located at distances $d < 10$\,kpc, we employed a different parallax cut. The parallax cut is dependent on distance, $d_{\rm in}$, as: $1/(2 d_{\rm in})< \varpi < 1.0/d_{\rm in}$ and changes as we are probing a distance range between 1.0 and 10\,kpc in increments of $d_{\rm in}$ of 1\,kpc up to 5 kpc, and then 2~kpc up to 9 kpc. In addition, we applied a brighter magnitude cut for 17, 18 and 19~mag for $d_{\rm in}=1$, 2 and 3 kpc cases, respectively, to use only the stars with better parallax accuracy.
A dSph in this distance range, whether completely disrupted or not, will likely only be recovered as a moving group with no discernible spatial overdensity, due to its large angular extent. However, we found that this does not penalise the XD search for a dSph. Hence, XD was applied to both proper motion and a spatial distribution as described above.

\section{Gaia DR2 search for a dwarf galaxy}
\label{sec:res}

\begin{figure*}
	\centering
	\includegraphics[width=\hsize]{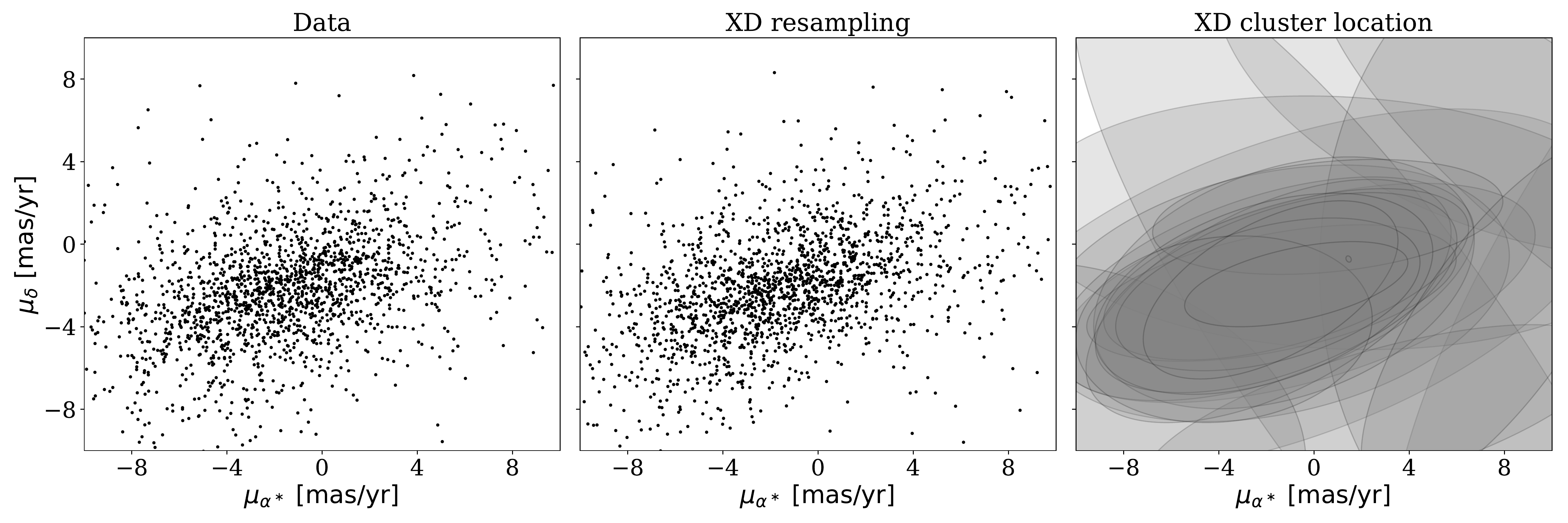}
	\caption{A detection of an open cluster -- NGC~7438 -- in the field of FHES J2304.0+5406. The left panel shows the data after filtering with $d_{\rm in}=1$~kpc, i.e.\ $0.5 < \varpi(\mathrm{mas}) < 1$. The right panel shows the detected Gaussian models with the XD method. The middle panel presents the distribution sampled from the detected models. The open cluster is marked by a small circle around $(\mu_{\alpha^*}, \mu_{\delta})=(1.45,-0.72)$~mas yr$^{-1}$ in the right panel. 
} 
	\label{fig:source8d1}
\end{figure*}

We applied XD to the samples of stars with $d<10$~kpc and $d>10$\,kpc, filtered as mentioned above, in the fields of 3FGL J$2212.5+0703$,  3FGL J$1924.8+1034$, FHES J$1501.0−6310$, FHES J$1723.5−0501$, FHES J$1741.6−3917$, FHES J$2129.9+5833$, FHES J$2208.4+6443$ and FHES J$2304.0+5406$. The intensive XD search undertaken after a careful visual inspection of all the fields using {\tt TOPCAT} \citep{Taylor2005ASPC..347...29T} yielded no evidence for a dSph in any of these fields. In the next section, we provide our conservative detection limits for each field. In this section, we provide examples of some interesting cases that demonstrate the validity of our approach.

While we found no evidence for new dSphs in these \Fermi-LAT fields, we did find overdensities of stars associated with known star clusters, demonstrating that the XD method can successfully find phase space structures in the \Gaia\ DR2 data. Fig.~\ref{fig:source8d1} shows that XD can be used to successfully detect faint stars in the outskirts of NGC~7438 ($l=106\fdg7$, $b=-4\fdg90$) which is at a distance of $d=1$ kpc \citep{Dias+2002A&A...389..871D}.
Our detected proper motion is $(\mu_{\alpha^*}, \mu_{\delta})=(1.45,-0.72)$~mas~yr$^{-1}$.

We also noticed that there is part of a globular cluster, NGC~6366, in the field of FHES J1723.5$-$0501. These stars are also clearly identified with the XD, and we found the proper motion of $(\mu_{\alpha^*}, \mu_{\delta})=(-0.35, -5.14)$~mas~yr$^{-1}$ for NGC~6366. The XD identified Ruprecht 112 ($d=1.76$~kpc) in the filed of FHES J1501.0$-$6310, when we set $d_{\rm in}=2$~kpc and also $d_{\rm in}=3$~kpc. Because fainter stars with lower parallax accuracy tend to be found at a higher distance \citep[e.g.][]{Luri+2018arXiv180409376L}, our method detects the stellar system when we use a larger $d_{\rm in}$ than the distance of the stellar system. We found $(\mu_{\alpha^*}, \mu_{\delta})=(-4.43, -4.23)$~mas~yr$^{-1}$ for Ruprecht 112. Finally, Trumpler~29 ($d=0.76$~kpc) is identified in the FHES J1741.6$-$43917 field. We obtain a proper motion of $(\mu_{\alpha^*}, \mu_{\delta})=(0.49, -2.30)$~mas~yr$^{-1}$ for Trumpler~29. All the above distances to the open clusters are from the catalogue of \citet{Dias+2002A&A...389..871D}. Note that our proper motions identified with the XD are different from those in \citet{Dias+2002A&A...389..871D}. This is likely due to the improved astrometric accuracy of the \Gaia\ data; we will explore this further in future work. 

\section{Mock data analysis: detection limits for dwarfs in Gaia DR2
}\label{sec:mock}

In this section, we estimate the detection limits for the XD method applied to the \Gaia\ DR2 data. We first estimate our detection limit using mock data as described in Section~\ref{sec:meth} for one of our primary \Fermi-LAT fields, 3FGL J2212.5+0703. We then place the mock dSph model with the estimated upper limit into the \Gaia\ data of each field and confirm that the mock dSph can be recovered in all of our fields up to a distance of $d=20$~kpc.

To estimate the detection limit in the 3FGL J2212.5+0703 field, we set up mock data as described in Section~\ref{sec:meth}. We placed a mock dSph with $M_*=10^3-10^4$~M$_{\sun}$, $r_h=20-100$~pc and $V_{\rm gal}=20-100$ km~s$^{-1}$ at a distance of $d=1-30$\,kpc. As in Section~\ref{sec:meth}, the field-star catalogues were made with {\tt Galaxia}, applying \Gaia\ DR2 uncertainties. Note that our goal is to place a conservative upper limit on the detection of a dSph in \Gaia\ DR2. The detection limits are sensitive to the parameters mentioned above, and precise evaluation would require more sophisticated mock data and statistical analyses that are beyond the scope of this paper. As we shall show, however, the conservative limit is sufficient for our calculation of the implications of our results for DM annihilation in Section \ref{sec:dm-disc}.

\begin{figure*}
	\centering
	\includegraphics[width=\textwidth]{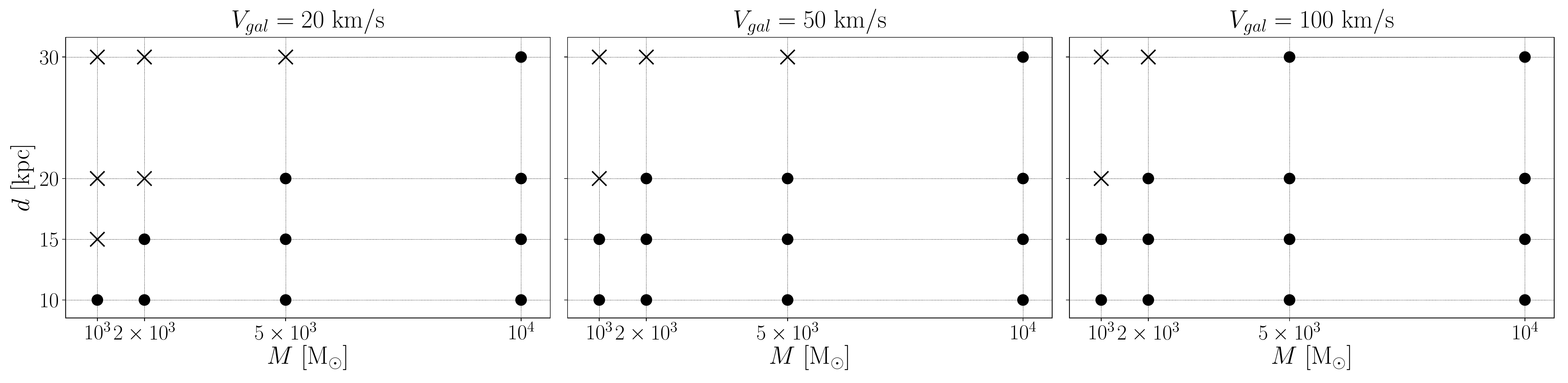}
	\caption{The detectability of a mock dSph galaxy as a function of its mass, distance and velocity using the XD method applied to the 3FGL J2212.5+0703 field. The left, middle and right panels show results for a velocity of 20, 50 and 100~km~s$^{-1}$, respectively. Solid circles (crosses) represent the parameters for which a dSph can (cannot) be found in the mock data with the XD.
} 
	\label{fig:coarse_search}
\end{figure*}

We found that $r_h$ is not a critical parameter for the detection limit, while our detection is sensitive to $V_{\rm gal}$. Hence, in the panels of Fig.~\ref{fig:coarse_search} we provide our detection limit parameter survey results for three different $V_{\rm gal}$ for the 3FGL J2212.5+0703 field. In all cases, the XD method reliably recovers a dSph of $M_* > 10^4$~M$_{\sun}$ at $d<20$~kpc.

 Based on this result, we now derive a conservative detection threshold for our other fields. For this, we set up mock dSphs with $M_*=10^4$~M$_{\sun}$, $r_h=50$~pc, $V_{\rm gal}=20$ km~s$^{-1}$ and $d=1-20$\,kpc for each field, as above, taking into account the different dust extinction in each field. (For this we use the {\tt SNAPDRAGONS} code that uses the same extinction model as {\tt Galaxia}.) For the fields 3FGL J2212.5+0703, FHES J1723.5$-$0501, FHES J2129.9+5833 and FHES J2208.4+6443, the XD method with $K=25$~components recovered the mock dSph up to 20~kpc. In the 3FGL J1924.8+1034, FHES J1501.0$-$6310 and FHES J2304.0+5406 fields, the dSph at $d=15$ and 20\,kpc required us to use $K=50$ components. In the densest field, FHES J1741.6$-$3917, the mock dSphs at $d>10$~kpc required $K=100$ components to be detected. In our search for dSphs in Section~\ref{sec:res}, we used the number of components, $K$, as determined from these mock data tests, to search for dSphs in each field. From the above analysis, we place a conservative upper limit on our detection threshold of $M_* > 10^4$~M$_{\sun}$ for any dSph galaxy along the line of sight to our sample of unassociated, extended, \Fermi-LAT within $d=20$~kpc.

\section{Dark matter interpretation of the \Fermi-LAT unassociated, extended sources}
\label{sec:dm-disc}

It was pointed out that the gamma-ray signatures from our primary two \Fermi-LAT unassociated sources 3FGL J2212.5+0703 and 3FGL J1924+1034 were consistent with DM annihilation into a $b\bar b$ final state with masses around tens of GeV \citep{Bertoni+16,Xia+17}. Since the mass of and distances to these objects were unknown, degeneracies between them and the annihilation cross section remained. 

For 3FGL J2212.5+0703, the radius that contains 68\% of the total gamma-ray photons is found to be $\sigma = 0\fdg25$ \citep{Bertoni+16}; for 3FGL J1924+1034 it is $\sigma = 0\fdg15$. If the density profile of the DM subhalo is well approximated by the Navarro-Frenk-White profile, characterized by the scale radius $r_s$ and characteristic density $\rho_s$ 
\citep{1997ApJ...490..493N}, $\sim$70\% of the total annihilation happens within $0.5r_s$. Therefore, we assume $\theta_s \equiv r_s/d \simeq 0\fdg5$ ($0\fdg3$) for 3FGL J2212.5+0703 (3FGL J1924+1034). This is based on the assumption that the tidal stripping is not too strong such that the tidal truncation radius is still larger than the scale radius, $r_t > r_s$, which is shown to be the case below.

\begin{figure*}
	\begin{center}
		\includegraphics[width=7.5cm]{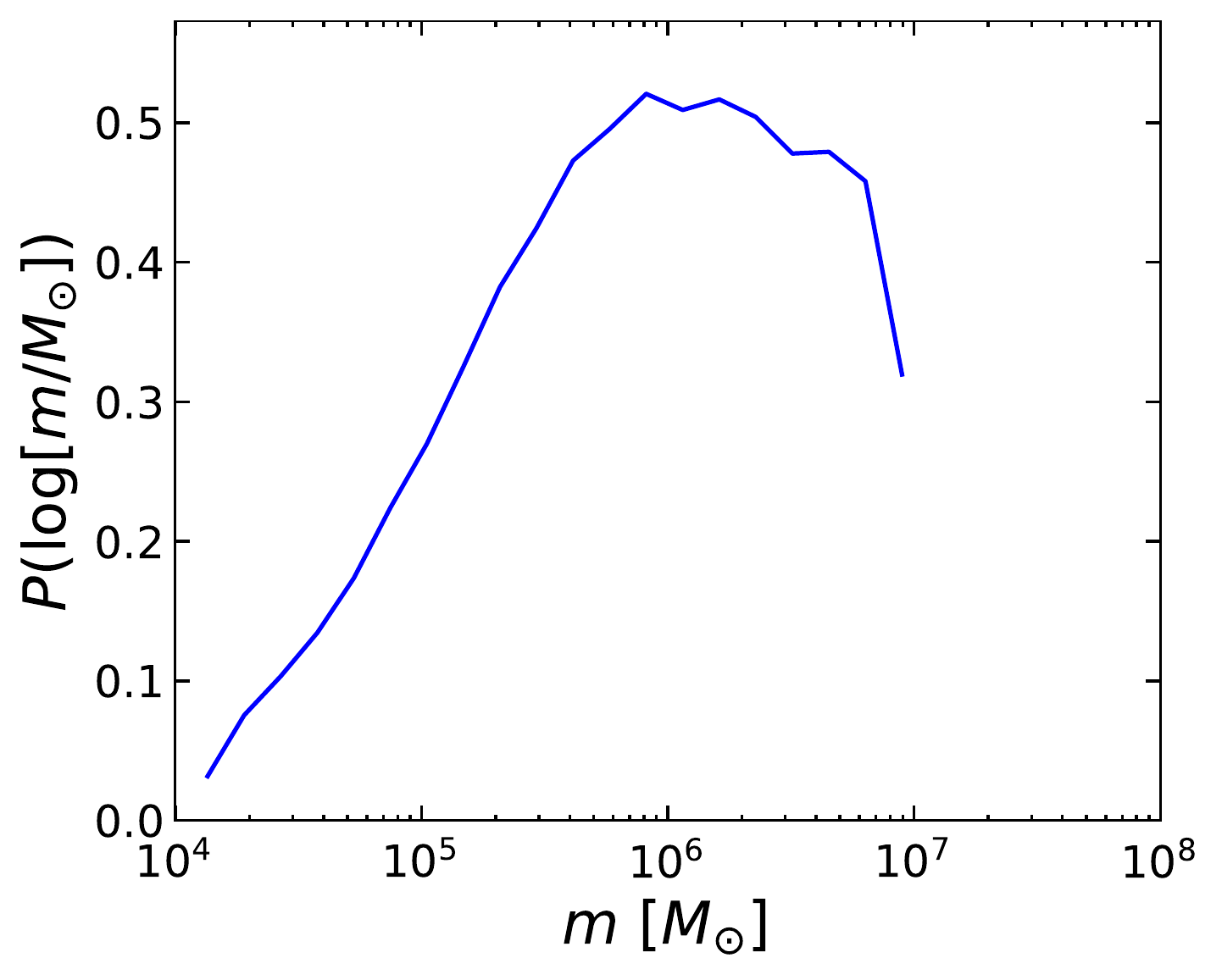}
		\includegraphics[width=7.5cm]{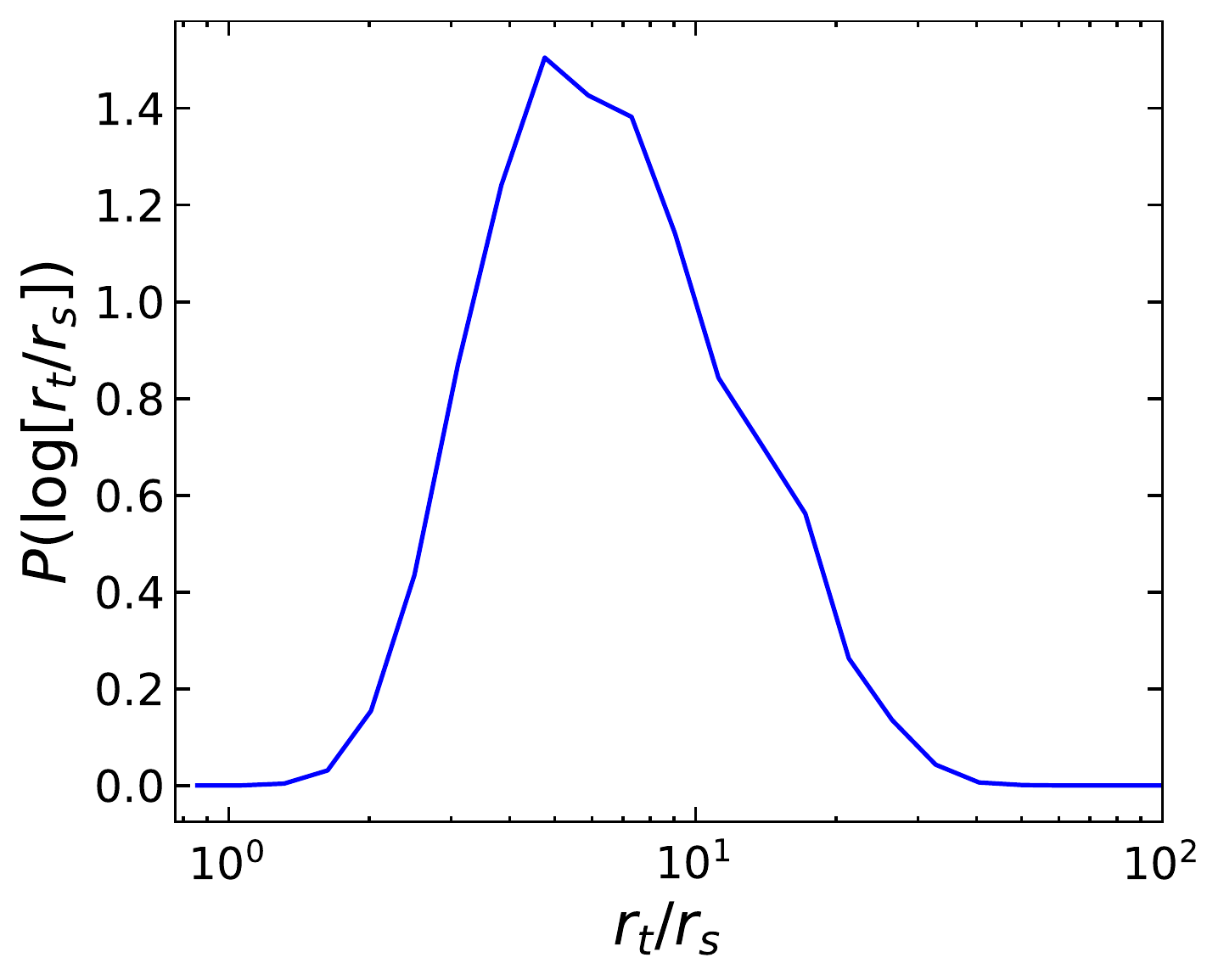}
		\includegraphics[width=7.5cm]{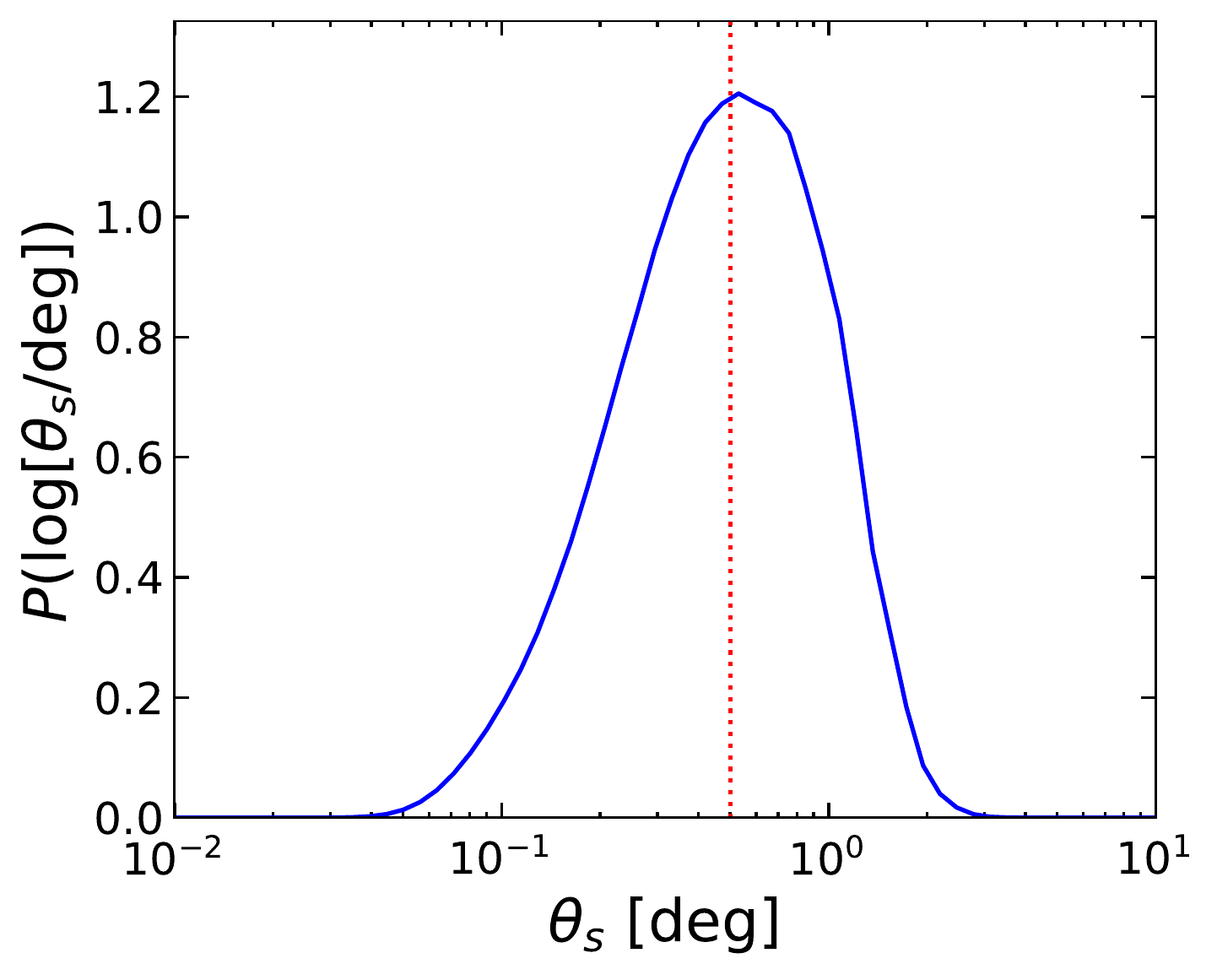}
		\includegraphics[width=7.5cm]{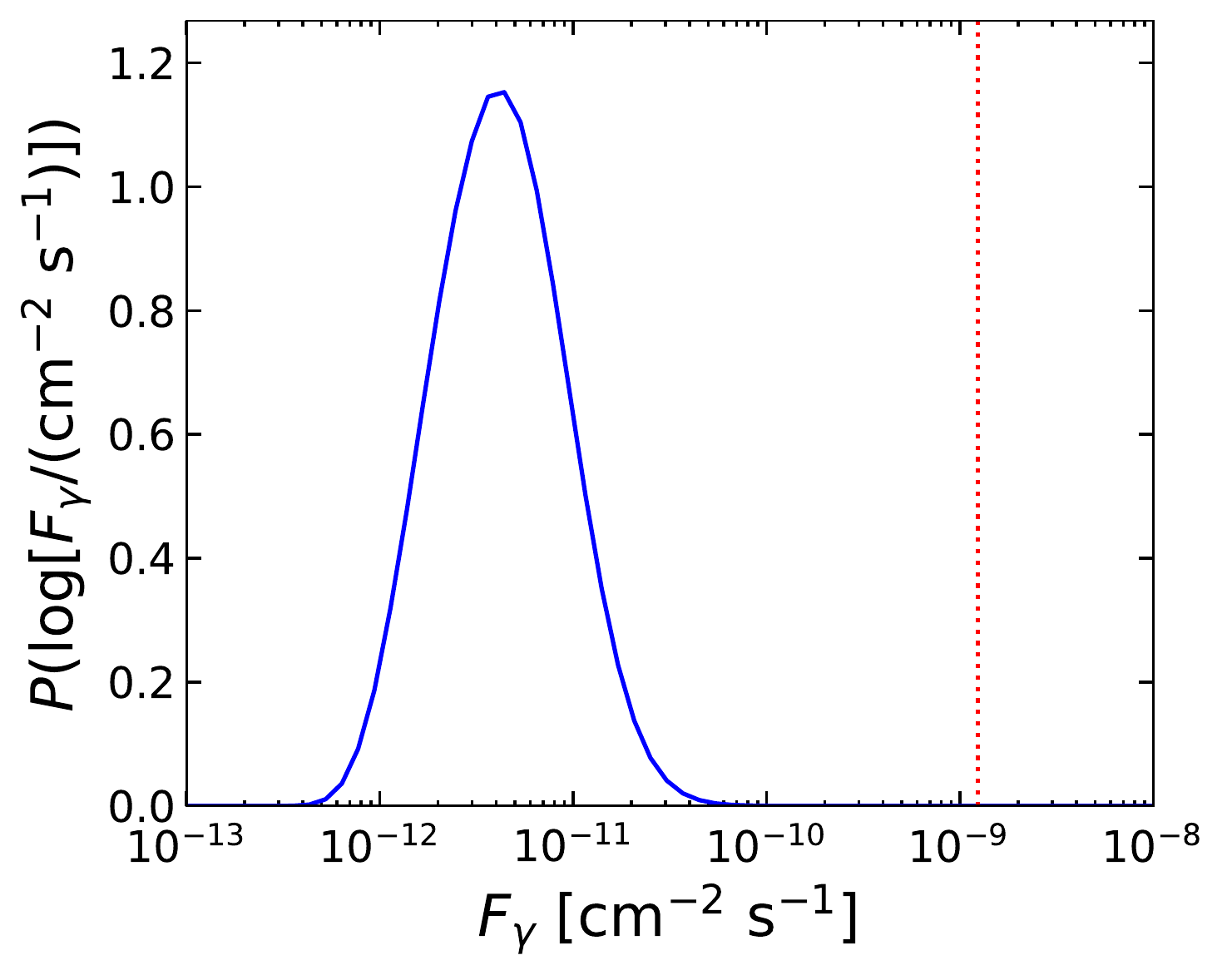}
		\caption{PDFs of post-infall subhalo mass (top left), ratio of tidal truncation radius and scale radius, $r_t / r_s$ (top right), the angle corresponding to the scale radius of the subhalo (bottom left), and the gamma-ray flux, $F_{\gamma}$, assuming $m_\chi = 25$~GeV and $\langle\sigma v\rangle = 2\times 10^{-26}$~cm$^3$~s$^{-1}$ (bottom right). In this example, the pre-infall halo mass is $M_{200} = 10^7M_{\sun}$ and the distance to the subhalo is $d = 10$\,kpc. The scale radius and flux inferred from the Fermi-LAT gamma-ray distribution for 3FGL J2212.5+0703 ($\theta_s = 0\fdg5$) are shown as vertical red dotted lines on the middle and bottom panels for comparison.}
		\label{fig:PDF_nodet}
	\end{center}
\end{figure*}

To constrain the possibility that the \Fermi-LAT sources we study in this paper owe to DM annihilation, we need to obtain a robust upper bound on the pre-infall halo mass, $M_{200}$, of dSphs with $M_* < 10^4$\,M$_\odot$ and $d < 20$\,kpc that lie below our detection threshold. This is because, at a fixed observed gamma-ray flux, the most massive, dense, halos minimise the inferred annihilation cross section. This then maximises the chance of consistency with the null detection from the known dSphs with $d>20$\,kpc.

Unfortunately, a satellite galaxy with $M_{*}<10^4$~M$_{\sun}$ is likely to have lost any tight relation between $M_*$ and its pre-infall $M_{200}$ \citep{2013MNRAS.433.2749G,2015NatCo...6E7599U,2017MNRAS.467.2019R}. However, we can estimate an upper bound on $M_{200}$ by comparison with the surviving `classical' dSphs, and via cosmological simulations of Milky Way-mass halos \citep[e.g.][]{2007ApJ...667..859D}. For the former, estimates of $M_{200}$ for classical dSphs like Draco and Ursa Minor with purely old stellar populations lie in the range $M_{200} = (2$--$5)\times 10^9$\,M$_\odot$ (Read \& Erkal 2018, in prep.). However, to be consistent with our detection threshold, these `classical' dSphs would have to lose over 90\% of their stellar mass \citep{2012AJ....144....4M}. The tidal stripping and shocking required to achieve this would lower their central dark matter densities by up to a factor of $\sim 10$ \citep[e.g.][]{2006MNRAS.367..387R}, significantly reducing the gamma-ray flux from DM annihilation. Indeed, a search for surviving subhalos in the {\tt Via Lactea II} simulation \citep{2007ApJ...667..859D} yielded only one subhalo with a pre-infall mass $M_{200} > 10^9$\,M$_\odot$ inside $d < 20$\,kpc, a subhalo with $M_{200} = 1.1 \times 10^9$\,M$_\odot$. Given the above arguments, we place a conservative upper bound on the  pre-infall halo mass of $M_{200} < 10^9$\,M$_\odot$ for surviving dSphs with $M_* < 10^4$\,M$_\odot$ and $d < 20$\,kpc. We consider this to be a strict upper bound since it does not account for subhalo depletion by the Milky Way disc \citep[e.g.][]{2010ApJ...709.1138D}.

Armed with an upper bound $M_{200}$, we now consider the effects of tidal mass loss due to the orbit of this subhalo around the Milky Way. We model this by using the analytic prescription in \citet{2015PhRvD..92l3508B} and \citet{2018arXiv180307691H}, assuming a $10^{12}M_{\sun}$ host halo for the Milky Way.

We first consider pre-infall subhalo masses of (0.8--$1.2)\times 10^7 M_{\sun}$ in order to model an example case, which is well below the upper limit from our non-detection of a dSph, but is chosen to illustrate our methodology. For a reference value of distance, we adopt $d = 10$~kpc.
In Fig.~\ref{fig:PDF_nodet}, we show the probability density function (PDF) of the subhalo mass after tidal stripping, $P(m)$ (top left),  and of the ratio of tidal truncation radius and the scale radius, $r_t / r_s$ (top right).
Although the subhalo can lose significant fraction of its mass due to the tidal effect, its truncation radius $r_t$ is kept larger than the scale radius $r_s$.
This is because we also take into account evolution of $r_s$ and $\rho_s$ \citep{2010MNRAS.406.1290P}, which compensates that of $r_t$. In the bottom left panel of Fig.~\ref{fig:PDF_nodet}, we show PDF of the angle size corresponding to the scale radius, $P(\theta_s)$ (middle).
We see that for this case, the gamma-ray source extension of $\sigma = 0\fdg25$ (3FGL J2212.5+0703; vertical red line) can be consistent with a pre-infall mass of $M_{200} = 10^7M_{\sun}$.

The gamma-ray flux from DM annihilation (assuming that DM is made of Majorana fermions) is calculated as:
\begin{equation}
F_\gamma = \frac{\langle\sigma v\rangle N_{\gamma,{\rm ann}}}{2 m_\chi^2}\frac{1}{4\pi d^2}\int dV \rho_{\chi}^2(r),
\label{eq:flux}
\end{equation}
where $m_\chi$ is the DM particle mass, $\langle\sigma v\rangle$ is the annihilation cross section, $N_{\gamma,{\rm ann}}$ is the number of gamma-ray photons emitted per annihilation, and $\rho_\chi(r)$ is the DM density profile of the subhalo.\footnote{Strictly speaking, Eq.~(\ref{eq:flux}) is valid only if the source extension is small, $\theta_s \ll 1$.  For smaller distances or larger masses, one has to implement the line-of-sight integral to compute the gamma-ray intensity and integrate it over the solid angle to obtain the flux.  However, the {\it observed} extensions of the \Fermi~unassociated sources are at most $\mathcal O(1)\degr$, and hence, in order to explain the observed features in terms of DM annihilation, using Eq.~(\ref{eq:flux}) instead of the more accurate line-of-sight integral is well justified.}
For a representative calculation, we adopt $m_\chi = 25$~GeV, $\langle \sigma v\rangle = 2\times 10^{-26}$~cm$^3$~s$^{-1}$ \citep[e.g.,][]{2012PhRvD..86b3506S}, and $N_{\gamma,{\rm ann}} = 3.7$ that is for $E_{\gamma}>1$~GeV in the case of annihilation into a $b\bar b$ final state.
The bottom panel of Fig.~\ref{fig:PDF_nodet} shows the PDF of the gamma-ray flux $F_\gamma$ corresponding to this example case of $M_{200} = 10^7M_{\sun}$ and $d = 10$~kpc.
It shows that, for this case in which one can explain the source extension well, the mean gamma-ray flux above 1~GeV will be $\sim$ $4\times 10^{-12}$~cm$^{-2}$~s$^{-1}$, which is smaller than the observed flux of $1.24\times 10^{-9}$~cm$^{-2}$~s$^{-1}$ for 3FGL J2212.5+0703 \citep{Bertoni+16} by more than two orders of magnitude.
This means that in order to explain the gamma-ray signal from 3FGL J2212.5+0703 in terms of DM annihilation, with a subhalo of this mass and distance, the annihilation cross section needs to be larger by more than two orders of magnitude than the canonical value ($\langle\sigma v\rangle = 2\times 10^{-26}$~cm$^3$~s$^{-1}$) adopted here as well as in \citet{Bertoni+16}.
Such a large cross section has already been excluded from the analysis of the known dSphs in the Milky Way \citep{2017ApJ...834..110A} at high significance, even when systematic uncertainties are taken into account~\citep{2018arXiv180305508C}. This illustrates how we can combine our \Gaia\ DR2 dSph detection limits with DM annihilation constraints from the known Milky Way dSphs to constrain or rule out a DM annihilation interpretation of the unassociated \Fermi-LAT sources.

\begin{figure}
\begin{center}
\includegraphics[width=7.5cm]{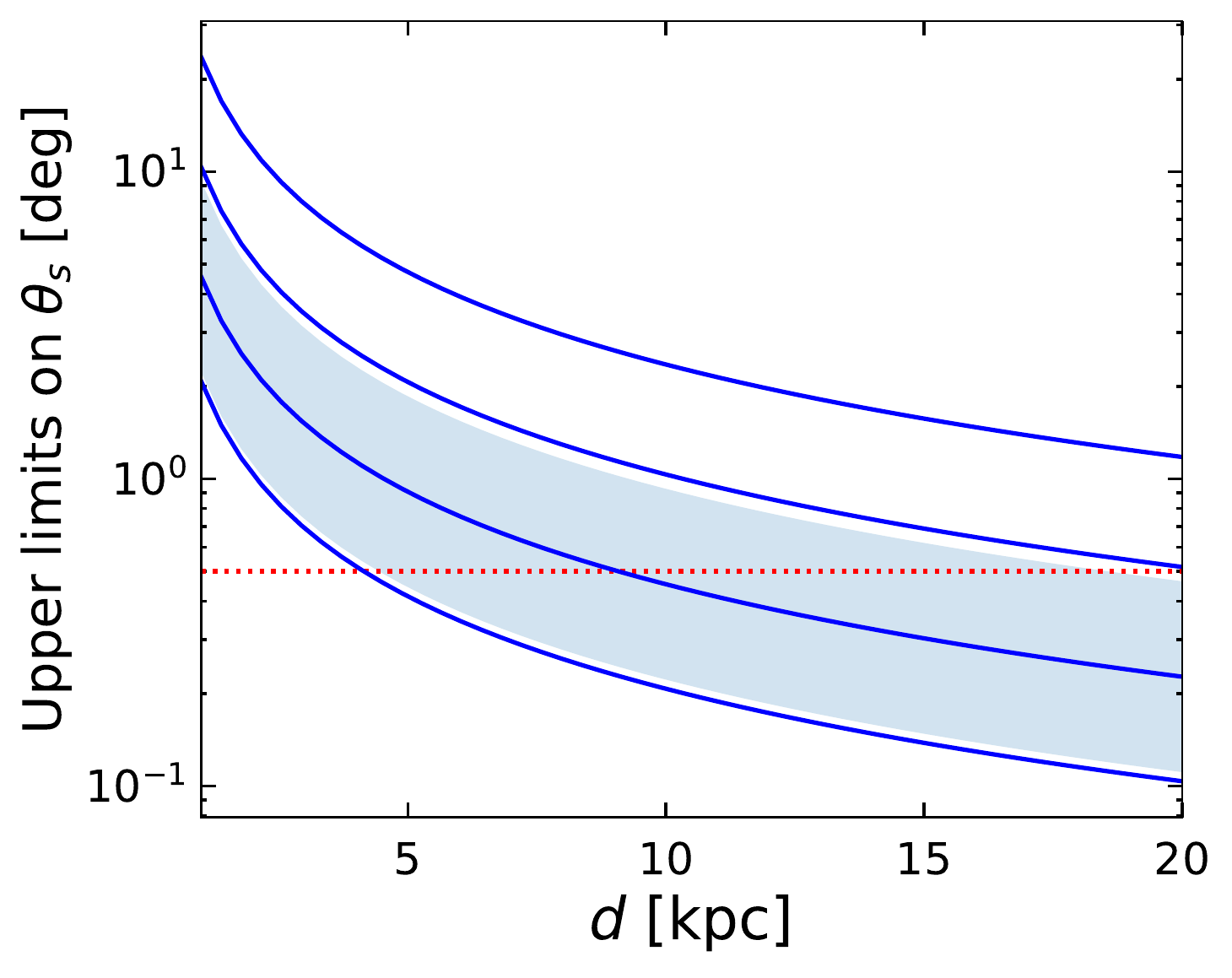}
\includegraphics[width=7.5cm]{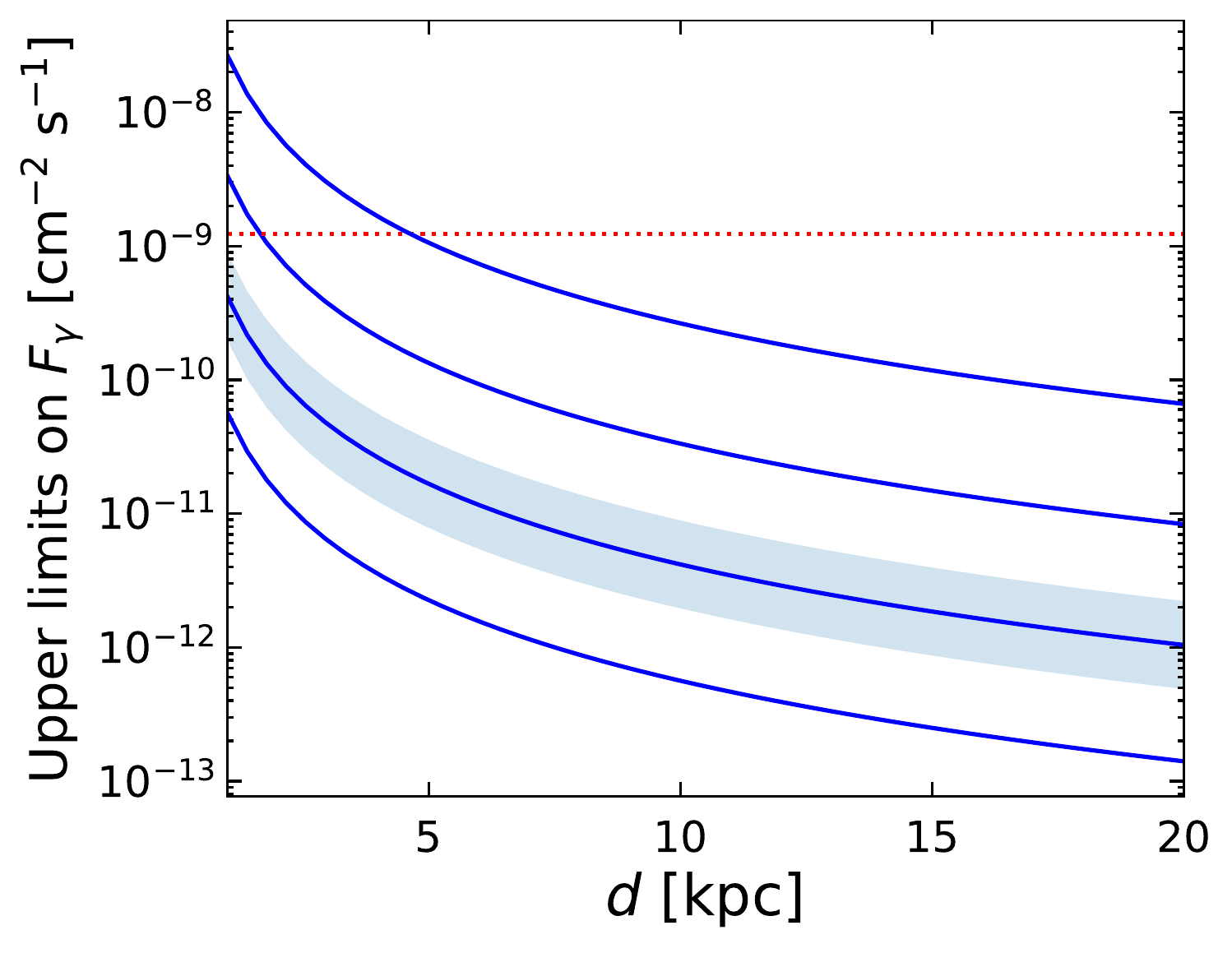}
\caption{Upper limits on the angular size corresponding to the scale radius $\theta_s$ (top) and gamma-ray flux (bottom) from DM annihilation $F_\gamma$ as a function of distance $d$. From top to bottom, the blue solid lines correspond to pre-infall halo masses: $M_{200} = 10^9$, $10^8$, and $10^6 M_{\sun}$, respectively. For the $M_{200} = 10^7$\,M$_\odot$ subhalo, we show the $1\sigma$ scatter in our theoretical modelling as a light blue band. This scatter owes primarily to the uncertain tidal mass loss history of the subhalo. The horizontal red dotted lines show the measured values from the \Fermi-LATdata analysis for 3FGL J2212.5+0703.} 
\label{fig:limits}
\end{center}
\end{figure}

In Fig.~\ref{fig:limits}, we show the upper limits on the angular extension $\theta_s$ (top panel) and on the gamma-ray flux (bottom panel) as a function of the dSph distance for the subhalo masses of $10^6$, $10^7$, $10^8$ and $10^9 M_{\sun}$ (solid blue lines, from bottom to top, respectively). (Recall that the upper limit on pre-infall mass is $M_{200} = 10^9 M_{\sun}$.) For the case of $M_{200} = 10^7M_{\sun}$, we also show the $1\sigma$ uncertainty in the modelling that owes primarily to the uncertain history of tidal mass stripping (blue band). 

We note that we also checked these results using a simple approach that does not account for tidal stripping and is based on the concentration parameterisation from~\citet{Moline:2016pbm}.
We find a good agreement between the two different approaches within our quoted uncertainties. The differences come from the fact that small changes of internal profiles such as $r_s$ and $\rho_s$ \citep[e.g.,][]{2010MNRAS.406.1290P} have been taken into account for the former approach based on \citet{2015PhRvD..92l3508B} and \citet{2018arXiv180307691H}, but not in the latter. 

\begin{figure}
\begin{center}
\includegraphics[width=7.5cm]{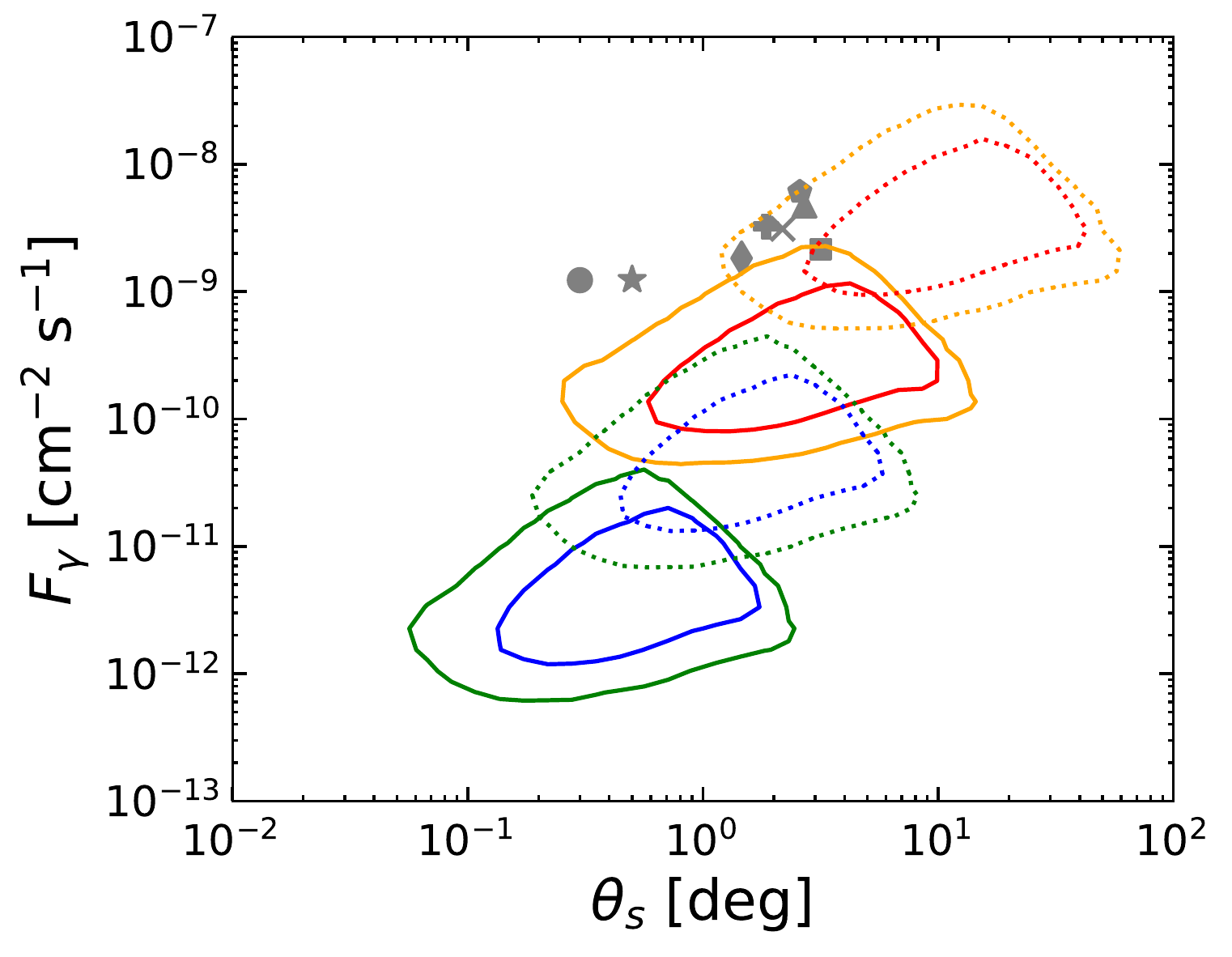}
\caption{$1\sigma$ and $2\sigma$ regions of the joint PDF, $P(\theta_s, F_\gamma)$, for subhalos with a pre-infall mass of $M_{200} = 10^7M_{\sun}$ at $d=10$~kpc (lower solid), $10^9M_{\sun}$ at 10~kpc (upper solid), $10^7M_{\sun}$ at 3~kpc (lower dotted), and $10^9M_{\sun}$ at 3~kpc (upper dotted). Measured values for the eight \Fermi~unassociated sources are shown for comparison: 3FGL J2212.5+0703 (star), 3FGL J1924.8$-$1034 (circle), FHES J1501.0$-$6310 (pentagon), FHES J1723.5$-$0501 (diamond), FHES J1741.6$-$3917 (square), FHES J2129.9+5833 (cross), FHES J2208.4+6443 (plus), and FHES J2304.0+5406 (square).}
\label{fig:thetas_flux}
\end{center}
\end{figure}

Fig.~\ref{fig:thetas_flux} shows the $1\sigma$ and $2\sigma$ contours in the plane, $(\theta_s,F_{\gamma})$, for pre-infall masses $M_{200} = 10^7M_{\sun}$ and $10^9M_{\sun}$ and distances $d = 3$~kpc and 10~kpc.
For the cases of 3FGL J2212.5+0703 (star; \citealt{Bertoni+16}) and 3FGL J1924.8-1034 (circle; \citealt{Xia+17}), the measured signals are incompatible with a DM annihilation interpretation, even for halos at the pre-infall halo mass upper bound of $M_{200} = 10^9$\,M$_\odot$.
We note that recently \citet{2018arXiv180408035F} reported that the gamma-ray signals from both 3FGL J2212.5+0703 and 3FGL J1924.8-1034 are better explained by two point sources rather than a single extended source. Our results here are consistent with this finding, excluding the hypothesis of DM annihilation from a subhalo within $d < 20$\,kpc.

\citet{2018arXiv180408035F} found six additional unassociated extended sources. In this work, we analysed all of them and found no signature of a dSph galaxy in any of these fields.
Among them, FHES J1723.5$-$0501, FHES J1741.6$-$3917, and FHES J2304.0+5406 were found to have an energy spectrum harder than $E^{-2}$, as is expected from DM annihilation.
In Fig.~\ref{fig:thetas_flux} we also show measured values of $(\theta_s,F_{\gamma})$ for these six \Fermi-LAT unassociated sources.
As can be seen, none of these unassociated sources is compatible with DM annihilation within our $2\sigma$ contours, unless the pre-infall mass of the subhalo is close to the upper limit, $10^9M_{\sun}$, and it is located around $d=3$~kpc. Such a massive pre-infall halo is very unlikely to be found so close to the Sun -- especially when accounting for subhalo depletion by the disc (see the discussion on this, above).
Since the annihilation cross section $\langle \sigma v \rangle = 2\times 10^{-26}~\mathrm{cm^3~s^{-1}}$ adopted here is already in tension with other data analyses \citep[e.g.,][]{2017ApJ...834..110A} for 25~GeV WIMPs, possibilities with lighter ($\lesssim 10^9M_{\sun}$) halos are excluded.

Finally, we discuss a few caveats, all of which we believe make our conclusion, above, stronger. Firstly, the discussions up to this point have been based on the assumption that the subhalo only experienced tidal stripping due to gravitational potential of the spherical host halo. However, subhalos orbiting within $d < 20$\,kpc are likely on eccentric orbits and will additionally experience tidal shocks that can lower their central density by up to a factor of $\sim 10$ \citep[e.g.][]{2006MNRAS.367..387R}. Furthermore, we have not taken into account the effect of the Milky Way disc that depletes the number of substructures within 20--30\,kpc by a factor of $\sim 2$ \citep{2010ApJ...709.1138D}. Including such effects will lower the gamma-ray flux for a given pre-infall $M_{200}$ requiring us to further increase the annihilation cross section, further exasperating the tension with the constraints from the known Milky Way dSphs. Finally, we have only considered subhalos within $d < 20$\,kpc. Lighter subhalos at larger distances than this would be too faint to be consistent with the observed fluxes unless they have large annihilation cross sections at odds with the constraints from the known dSphs in the Milky Way. Massive subhalos with $M_{200} > 10^9$\,M$_\odot$ at larger distances would have a readily detectable stellar counterpart, and then a DM annihilation signal from the known dSphs in the Milky Way should be already detected (see the discussion, above). Hence, this can also be excluded. We conclude that none of the unassociated, extended, \Fermi-LAT sources studied here can have a DM annihilation origin.

\section{Summary}
\label{sec:sum}

Using the new second data release of the \Gaia\ mission, we have made the first attempt to find a dSph towards eight \Fermi-LAT extended unassociated source fields. Our goal was to link the gamma-ray emission, already detected by \Fermi-LAT, to a possible optical counterpart within the framework of annihilating DM. \Gaia's superb astrometric accuracy provides a new window for searching for dSphs in the inner Galactic halo ($d<20$~kpc) based on the proper motion and parallax of stars \citep{Antoja+15}. We applied an advanced statistical method, the Extreme-Deconvolution \citep[XD,][]{Bovy+11} Gaussian Mixture Model \citep[XDGMM,][]{Holoien+17}, to properly take into account the uncertainties and correlations in proper motion space. Unfortunately, we found no detection indicative of a signature of a dSph galaxy in any of these fields placed within the \Gaia\ data. We then estimated the detection limits for a dSph galaxy by applying XD to mock data. We obtained a conservative limit on the stellar mass of any undetected dSph of $M_* < 10^4$~M$_{\sun}$ for $d < 20$\,kpc. We showed that this corresponds to an upper limit on the pre-infall halo mass of $M_{200} < 10^9$~M$_{\sun}$.

Using an analytical model of subhalo mass stripping that has been calibrated against numerical simulation results, combined with current limits on the DM annihilation cross section from known Milky Way dSphs, we estimated the gamma-ray flux and source size as a function of pre-infall halo mass for all eight \Fermi-LAT sources. We concluded that our model rejects the possibility of a DM annihilation scenario for the two sources: 3FGL J2212.5+0703 \citep{Bertoni+16} and 3FGL J1924.8-1034 \citep{Xia+17} if the pre-infall DM halo mass is less than $M_{200} < 10^9$\,M$_\odot$ at distance of $d < 20$\,kpc, as constrained by our work. If the subhalo is farther away than 20~kpc, then its DM halo has to be larger than $M_{200} = 10^9$\,M$_\odot$. Such a dSph would have a clear stellar counterpart that should have been detected in the existing photometric data in these fields. Furthermore, if there were a dSph at $d>20$ kpc with a mass similar to the dSphs already discovered, then we would expect to see also a DM annihilation signal from the known dSphs in the Milky Way. Hence, we conclude that a DM origin for these two sources is rejected.

We then applied our model and dSph constraints to the six \Fermi-LAT extended unassociated sources recently found in \citet{2018arXiv180408035F}. We concluded that these too are unlikely to have a DM origin. We could only explain them as coming from DM annihilation if they owe to a DM subhalo with a pre-infall halo mass $M_{200} = 10^9$\,M$_\odot$ and a distance less than $d=3$~kpc. We are unable to find any such subhalo in the {\tt Via Lactea II} cosmological simulation. When accounting for the unmodelled effects of tidal stripping and shocking by the Milky Way stellar disc, such subhalos become even rarer. We conclude, therefore, that none of the unassociated, extended, \Fermi-LAT sources found to date is likely to have a DM annihilation origin.

Our work represents the first attempt to search for stellar counterparts to \Fermi-LAT extended sources that could owe to DM annihilation. The XD method was able to find known star clusters in the fields studied, demonstrating its efficacy. But we were unable to find evidence for any new dSphs in these fields. In future work, we will perform a similar search for stellar counterparts along the line of sight to all \Fermi-LAT unassociated sources, including point-like objects.

\section*{Acknowledgement}
IC and DK acknowledge the support of the UK's Science and Technology Facilities Council (STFC Grant ST/K000977/1 and ST/N000811/1). IC is also grateful the STFC Doctoral Training Partnerships Grant (ST/N504488/1). IC thanks the LSSTC Data Science Fellowship Program, where their time as a Fellow has benefited this work. The work of SA was supported in part by JSPS KAKENHI Grants (17H04836, 18H04340, and 18H04578). JIR would like to acknowledge support from SNF grant PP00P2\_128540/1, STFC consolidated grant ST/M000990/1 and the MERAC foundation. CM acknowledges support from grant DGAPA/UNAM IG100115. This work has made use of data from the European Space Agency (ESA) mission \Gaia\ (\url{https://www.cosmos.esa.int/gaia}), processed by the \Gaia\ Data Processing and Analysis Consortium (DPAC, \url{https://www.cosmos.esa.int/web/gaia/dpac/consortium}). Funding for the DPAC has been provided by national institutions, in particular, the institutions participating in the \Gaia\ Multilateral Agreement. This research has made use of the VizieR catalogue access tool, CDS, Strasbourg, France. The original description of the VizieR service was
 published in A\&AS 143, 23. 
\bibliographystyle{mnras}
\bibliography{ms,ufs}
%\bibliography{ufs}
% if your bibtex file is called example.bib

% \begin{thebibliography}{99}
% \bibitem[\protect\citeauthoryear{Author}{2012}]{Author2012}
% Author A.~N., 2013, Journal of Improbable Astronomy, 1, 1
% \bibitem[\protect\citeauthoryear{Others}{2013}]{Others2013}
% Others S., 2012, Journal of Interesting Stuff, 17, 198
%\end{thebibliography}

%%%%%%%%%%%%%%%%%%%%%%%%%%%%%%%%%%%%%%%%%%%%%%%%%%

%%%%%%%%%%%%%%%%% APPENDICES %%%%%%%%%%%%%%%%%%%%%

% \appendix

% \section{Some extra material}

% it can be placed in an Appendix which appears after the list of references.

%%%%%%%%%%%%%%%%%%%%%%%%%%%%%%%%%%%%%%%%%%%%%%%%%%

% Don't change these lines
\bsp	% typesetting comment
\label{lastpage}
\end{document}